\baselineskip=15pt
\def\b{\beta}\def\d{\delta}\def\e{\epsilon}
\def\h{\theta}
\def\l{\lambda}\def\m{\mu}\def\n{\nu}\def
\p{\pi}\def\r{\rho}
\def\y{\eta}

\def\D{\Delta}
\def\O{\Omega}

\def\id{\equiv}\def\mo{{-1}}

\def\({\left(}\def\){\right)}\def\[{\left[}\def\]{\right]}

\def\mn{{\mu\nu}}

\def\pb{Poisson brackets }
\def\sch{Schwarzschild }\def\cc{coupling constant }\def\coo{coordinates }
\def\GR{general relativity }
\def\wrt{with respect to }

\def\section#1{\bigskip\noindent{\bf#1}\smallskip}

\def\PL#1{Phys.\ Lett.\ {\bf#1}}

\def\PR#1{Phys.\ Rev.\ {\bf#1}}\def\CQG#1{Class.\ Quantum Grav.\ {\bf#1}}
\def\NP#1{Nucl.\ Phys.\ {\bf#1}}

\def\JoP#1{J.\ Phys.\ {\bf#1}} \def\IJMP#1{Int.\ J. Mod.\ Phys.\ {\bf #1}}

\def\JHEP#1{JHEP\ {\bf#1}}

\def\arx#1{{\tt arXiv:#1}}

\def\ref#1{\medskip\everypar={\hangindent 2\parindent}#1}
\def\beginref{\begingroup
\bigskip
\centerline{\bf References}
\nobreak\noindent}
\def\endref{\par\endgroup}

{\nopagenumbers
\line{}
\vskip60pt
\centerline{\bf Snyder dynamics in a \sch spacetime}
\vskip60pt
\centerline{
{\bf S. Mignemi}$^*$\footnote{$^\ddagger$}{e-mail: smignemi@unica.it}
and {\bf R. \v Strajn}\footnote{$^\dagger$}{e-mail: ri.strajn1@studenti.unica.it}}
\vskip10pt
\centerline {Dipartimento di Matematica e Informatica, Universit\`a di Cagliari}
\centerline{viale Merello 92, 09123 Cagliari, Italy}
\smallskip
\centerline{$^*$and INFN, Sezione di Cagliari}
\vskip80pt
\centerline{\bf Abstract}
\medskip
\noindent We calculate the orbits of a particle in Schwarzschild spacetime, assuming that the
dynamics is governed by a Snyder symplectic structure. With this assumption, the perihelion
shift of the planets acquires an additional contribution with respect to the one predicted by
general relativity. Moreover, the equivalence principle is violated.
If one assumes that Snyder mechanics is valid also for macroscopic systems, these
results impose strong constraints on the value of the coupling parameter of the Snyder model.
\vskip10pt
{\noindent

}
\vskip80pt\
\vfil\eject}
\section{1. Introduction}
Noncommutative geometry is becoming a serious candidate to describe spacetime at Planck
scales, where quantum gravity effects are sensible. In particular it accounts for the
existence of a minimal measurable length, that seems to be a common outcome of different
quantum gravity theories.

Among the many possible versions of noncommutative geometry, a
special place is taken by its original formulation, proposed by Snyder [1],
since, contrary to many of its rivals, this model preserves the Lorentz invariance,
which is at the basis of the present understanding of physics.

Although the validity of noncommutative geometry is presumably limited to Planck-scale
physics, it may be interesting to investigate if its effects can extend to macroscopic
systems, where the classical limit holds, like for example the solar system.
In our point of view, this is not plausible, since noncommutative geometry is supposed to
hold only at scales where quantum gravity is effective, whereas extending
its validity much beyond this realm one risks to be faced with problems analogous to
the so-called soccer-ball problem of doubly special relativity [2], which shows
that paradoxical effects arise if one tries to apply deformed momentum relations
(analogous to those holding in Snyder mechanics) to macroscopic bodies.

This opinion is confirmed also by previous studies of planetary motion based on Snyder
dynamics [3], that when confronted with observations predict for the coupling constant
of the model a scale well below the Planck scale that would be expected on dimensional
grounds.

These estimates have however been obtained from a Newtonian theory, while the effect of
general relativity cannot certainly be neglected at these scales. For this reason
in the present paper we repeat the calculation of Snyder planetary orbits in a
relativistic setting. The results will partially confirm those of previous works [3],
since the corrections to relativistic dynamics due to Snyder mechanics will turn out to
be of the same order of magnitude
as the ones obtained in the Newtonian approximation, although numerically different.

\medskip
We recall that the Snyder model, in its classical limit, is based on the noncanonical
\pb [1]
$$\{x_\m,p_\n\}=\y_\mn+\b^2p_\m p_\n,\qquad\{x_\m,x_\n\}=\b^2J_\mn,\qquad\{p_\m,p_\n\}=0,
\eqno(1.1)$$
where $J_{\mn}=x_\n p_\n-x_\n p_\m$, $\y_\mn$ is the flat metric with signature $(-1,1,1,1)$
and $\b$ a coupling constant that is assumed to be of order one in Planck units.
In ordinary unities, this corresponds to $\b\sim\sqrt\hbar/cM_{Pl}\sim10^{-17}$(s/kg)$^{1/2}$.
The \pb (1.1) preserve the Lorentz invariance, but deform the action of translations on
spacetime [4]. Moreover, spacetime \coo satisfy nontrivial brackets, that are the classical
mechanics counterpart of spacetime noncommutativity.

The implications of the Snyder model have been studied from several points of views, either
in their classical or quantum aspects [5].
Also the generalization to spaces of constant curvature has been considered to some extent
[6].
However, in most cases the investigations have been limited to the nonrelativistic version
of the theory, essentially because the relativistic model poses several
technical and conceptual problems. To our knowledge a concrete example of relativistic
dynamics has only been considered in [7] in the case of the harmonic oscillator.

Our approach to the problem of planetary orbits will be rather conservative:
we write down the Hamilton equation
of a free particle in a \sch background, and assume that the only changes in the dynamics
are due to the Snyder noncanonical
symplectic structure (1.1). In particular, we shall choose the same Hamiltonian as in
general relativity, although the Snyder symmetries may allow for more general choices.

The \sch geodesics will be slightly deformed. In particular, a shift of the perihelion arises
in addition to that predicted by general relativity, whose sign is however opposite to the
one obtained from the calculation based on Newtonian gravity.

Another important outcome of our investigation is that the principle of equivalence is
broken in Snyder mechanics, since the corrections to the equation of the geodesics depend
on a parameter $\b^2m^2$, which is a function of the mass $m$ of the particle.
This effect is a consequence of the nontrivial dependence of the dynamics on the momenta
of the particles, and it also puts
strong limits on the value of the \cc $\b$ if the validity of Snyder mechanics
at planetary scale is assumed.

Of course, the limitation of the validity of Snyder mechanics to microscopic physics should
be justified. As mentioned before, this problem can be related to the soccer-ball problem of
doubly special relativity: in fact, in Snyder spacetime the summation rules for the momenta
must be nonlinear, since the translation invariance is deformed [4], and, following a reasoning
analogous to that of ref.\ [2], should be arranged in such a way that classical mechanics
holds at macroscopic scales.
A related argument, that has not been thoroughly investigated yet, is that passing from the
quantum-gravity regime to its classical limit some kind of decoherence should occur and hence
classical mechanics is recovered, as in the classical limit of quantum mechanics.
A discussion of this idea would however require a more definite theory of quantum gravity
than available a present.

\section{2. Particle motion in flat spacetime}
In order to set the formalism, we start by considering the free motion of a particle
in three-dimensional flat Snyder spacetime. We parametrize the spatial sections with polar
coordinates, defined in terms of cartesian \coo as
$$t=x_0=-x^0,\qquad\r=\sqrt{(x^1)^2+(x^2)^2},\qquad\h=\arctan{x^2\over x^1}.\eqno(2.1)$$
The corresponding momentum components read
$$p_t=p_0,\qquad p_\r={x^1p_1+x^2p_2\over\sqrt{(x^1)^2+(x^2)^2}},
\qquad p_\h\id J_{12}=x_1p_2-x_2p_1.\eqno(2.2)$$
With these definitions, the \pb for polar \coo in Snyder space following from (1.1) are
$$\{t,p_t\}=-1+\b^2p_t^2,\qquad\{\r,p_\r\}=1+\b^2\left(p_\r^2+{p_\h^2\over\r^2}\right),
\qquad\{\h,p_\h\}=1,$$
$$\{\r,\h\}=\b^2{p_\h\over\r},\qquad\{t,\r\}=\b^2(tp_\r-\r p_t),
\qquad\{t,\h\}=\b^2{tp_\h\over\r^2},$$
$$\{p_t,p_\r\}=-\b^2{p_tp_\h^2\over\r^3},\qquad\{p_t,p_\h\}=\{p_\r,p_\h\}=\{t,p_\h\}=
\{\r,p_\h\}=0,$$
$$\{t,p_\r\}=\b^2\(p_tp_\r+{tp_\h^2\over\r^3}\),\qquad\{\r,p_t\}=\b^2p_tp_\r,
\qquad\{\h,p_t\}=\b^2{p_t p_\h\over\r^2},\qquad\{\h,p_\r\}=\b^2{p_\r p_\h\over\r^2}.
\eqno(2.3)$$
Note that, contrary to the canonical case, the choice of polar \coo changes the symplectic
structure.

The Hamiltonian is chosen as in special relativity
$$H={\l\over2}\(-p_t^2+p_\r^2+{p_\h^2\over\r^2}+m^2\)=0,\eqno(2.4)$$
with $\l$ a Lagrange multiplier enforcing the mass shell constraint.
The choice of the Hamiltonian is not unique, but (2.4) seems to be the
most reasonable in this context.

The Hamilton equations derived form (2.3) and (2.4) are
$$\eqalignno{&\dot t=\l\D p_t,\qquad\dot\r=\l\D p_\r,\qquad\dot\h=\l\D {p_\h\over\r^2},&\cr
&\quad\dot p_t=0,\qquad\dot p_\r=\l\D {p_\h^2\over\r^3},\qquad\dot p_\h=0,&(2.5)}$$
with $\D=1-\b^2m^2$. Hence, as in special relativity, the momenta $p_\h$ and $p_t$ are constants
of the motion, that according to the standard notations we denote $ml$ and $E$ respectively.
They can be identified with the angular momentum and energy of the particle.
As in 1+1 dimensions [7] all the equations are identical to those of classical relativity,
except that they are multiplied by the common factor $\D$. Their solution can therefore be
obtained as in special relativity, after a redefinition of the proper time.

In particular one should choose a gauge by fixing the time variable, in order to eliminate the
Hamiltonian constraint (2.4) by means of the Dirac formalism.
However, if one is only interested in the equation of the orbits, it is not necessary to fix
the gauge since, like in special relativity,
$${d\r\over d\h}={\dot\r\over\dot\h}=\r^2{p_\r\over p_\h},\eqno(2.6)$$
does not depend on $\l$.
From the Hamiltonian constraint (2.4) follows that
$$p_\r=\sqrt{E^2-m^2\(1+{l^2\over\r^2}\)},\eqno(2.7)$$
and hence
$$\r'\id{d\r\over d\h}={\r\over l}\ \sqrt{\({E^2\over m^2}-1\)\r^2-l^2},\eqno(2.8)$$
which is solved by
$$\r={l\over\sqrt{E^2/m^2-1}}\ {1\over\cos(\h-\h_0)},\eqno(2.9)$$
that describes a straight line in polar coordinates and coincides with the
solution of classical special relativity.

\section{3. Particle motion in \sch spacetime}
We pass now to study the motion of a planet in the \sch spacetime with metric
$$ds^2=-A(\r)\,dt^2+A^\mo(\r)\,d\r^2+\r^2d\O^2,\eqno(3.1)$$
where
$$A(\r)=1-{2M\over\r}\eqno(3.2)$$
and $M$ is the mass of the sun. As in special relativity, due to the conservation
of the angular momentum, the problem can be reduced to 2+1 dimensions.

The Hamiltonian is chosen as in standard relativity,
$$H={\l\over2}\[-{p_t^2\over A}+Ap_\r^2+{p_\h^2\over\r^2}+m^2\]=0,\eqno(3.3)$$
where $m$ is the mass of the planet.

The field equations derived from (2.3) and (3.3) are
$$\dot t=\l\[p_t\(A^\mo-\b^2m^2-\b^2{M\over\r}\(p_\r^2+{p_t^2\over A^2}\)\)+
\b^2{Mtp_\r\over\r}\(p_\r^2+{p_t^2\over A^2}-2{p_\h^2\over\r^2}\)\],$$
$$\dot\r=\l\[A-\b^2m^2-{2\b^2Mp_\h^2\over\r^3}\]p_\r,\qquad
\dot\h=\l{p_\h\over\r^2}\[1-\b^2m^2-{\b^2M\over\r}\,\(p_\r^2+{p_t^2\over A^2}\)\],$$
$$\dot p_t=-\l\[{\b^2Mp_tp_\r\over\r^2}\(p_\r^2-{2p_\h^2\over\r^2}+{p_t^2\over A^2}\)\],
\qquad\qquad\dot p_\h=0,$$
$$\dot p_\r=\l\[(1-\b^2m^2){p_\h^2\over\r^3}-{M\over\r^2}\[\(p_\r^2+{p_t^2\over A^2}\)
\(1+\b^2\(p_\r^2+{p_\h^2\over\r^2}\)\)-2\b^2{p_\r^2p_\h^2\over\r}\]\].\eqno(3.4)$$

We are only interested in the equation of the orbits. To find it we proceed as in the
previous section.
While $p_\h$ is still a constant, $p_t$ is no longer conserved. Instead, one can check
that the quantity
$$E={p_t\over\sqrt{1+\b^2(-p_t^2+p_\r^2+p_\h^2/\r^2)}}\eqno(3.5)$$
is conserved and plays the role of the energy.
It follows that
$$p_t^2={E^2\over1+\b^2E^2}\[1+\b^2\(p_\r^2+p_\h^2/\r^2\)\].\eqno(3.6)$$
Moreover, (3.3) and (3.6) imply that
$$p_\r^2={E^2(1+\b^2m^2l^2/\r^2)-m^2(1+\b^2E^2)(1+l^2/\r^2)A\over(1+\b^2E^2)A^2-\b^2E^2}
\eqno(3.7)$$
where we have defined $l=p_\h/m$.

The equation of the orbits is conveniently written in terms of the variable $u=1/\r$ as
$${du\over d\h}=-{1\over\r^2}{\dot\r\over\dot\h}=-{A-\b^2m^2(1+2Ml^2u^3)
\over1-\b^2m^2-\b^2Mu\(p_\r^2+p_t^2/A^2\)}\ {p_\r\over ml}.\eqno(3.8)$$
Substituting in (3.8) the values of $p_\r$ and $p_t$ deduced from (3.6) and (3.7)
one can write down a differential equation for the single variable $u(\h)$.

The calculations are very involved, and the equation can only be solved perturbatively.
One can first expand in the Snyder parameter $\b^2m^2$ and then adopt the usual expansion
used in standard textbooks on \GR to solve for the \sch orbits.
To this end, it is useful to define the dimensionless quantities $v={l^2\over M}\,u$ and
$\e={M^2\over l^2}$. The parameter $\e$ is small for planetary orbits, and can be taken as
an expansion parameter.  We assume moreover that $\b^2m^2\ll\e$ since the Snyder
corrections are expected to be small \wrt those of general relativity.
Moreover, by the virial theorem, and the definition (3.5) of $E$,
$E^2-m^2\sim m^2(\e\,q+\b^2E^2)$, with $q$ a parameter of order unity.

The first-order expansion in both $\b^2m^2$ and $\e$ gives, after lengthy calculations,
$$v'^2=q+2v-v^2+2\e v^3+\b^2m^2\big[2v+4\e(qv+v^2)\big].\eqno(3.9)$$
It is convenient to take the derivative of this expression. One has
$$v''=1+\b^2m^2-v+\e[3v^2+\b^2m^2(2q+4v)].\eqno(3.10)$$
Expanding $v=v_0+\e v_1+\dots$, at zeroth order one obtains a Newtonian approximation of
the solution,
$$v_0=1+\b^2m^2+e\cos\h,\qquad\qquad e=1+{q\over\e}=1+{l^2(E^2-m^2)\over M^2m^2 },
\eqno(3.11)$$
while $v_1$ satisfies
$$v_1''+v_1=3+(10+2q)\b^2m^2+2(3+5\b^2m^2)e\cos\h+3e^2\cos^2\h,\eqno(3.12)$$
which is solved by
$$v_1=3\(1+{e^2\over2}\)+2\b^2m^2(5+q^2)+e(3+5\b^2m^2)\h\sin\h-{e^2\over2}\,\cos2\h.
\eqno(3.13)$$
The solution at first order is therefore
$$v\sim(1+\b^2m^2)+\e\[3\(1+{e^2\over2}\)+2\b^2m^2(5+q^2)\]+
e\cos\big[\big(1-\e(3+5\b^2m^2)\big)\h\big]-{\e\over2}\,e^2\cos2\h.$$

From this expression one can easily obtain the perihelion shift as
$$\d\h=2\p\e(3+5\b^2m^2)\sim{6\p M^2\over l^2}\(1+{5\over3}\b^2m^2\).$$
The first term is of course the one predicted by general relativity, while the second
depends on the mass of the planet.
This dependence is of course a consequence of the breaking of the equivalence
principle in Snyder mechanics.

In a Newtonian setting, the shift due to Snyder mechanics is given by
$\d\h=-2\p\b^2m^2M^2/l^2$ [3].
While the order of magnitude of the Snyder correction is the same as that obtained
from the relativistic model, its sign is opposite. Therefore, calculations based on
Newtonian mechanics are not much reliable in this context.
In any case, it has been shown [3] that for these
corrections to be compatible with the observed discrepancy of the perihelion shift of
Mercury from the predictions of general relativity, $\b$ must be less than $10^{-9}$
in Planck units. This estimate remains true in the relativistic case.

Another bound on the value of $\b$ can be obtained from the breaking of the
equivalence principle caused by the presence of terms proportional to $\b^2m^2$ in the
corrections to the geodesics motion.
Experimental data show that violation of the equivalence principle are less than one
part in $10^{12}$ [8].
It follows that $\b<10^{-26}$ in Planck units for planetary masses of order $10^{24}$kg
$=10^{32}$M$_{Pl}$.
This bound is even stronger than the previous one.

These results seem to indicate that if one assumes that Snyder mechanics holds
at scales compatible with the orbit of planets, the coupling constant $\b$ must be
less than its natural value of order 1 in Planck units by many orders of magnitude.

As discussed in the introduction, the most reasonable solution to this problem is that
Snyder mechanics be valid only at Planck scales, while at larger scales the dynamics
becomes classical, although the detailed mechanism of this transition has not been
figured out yet.

\beginref
\ref [1] H.S. Snyder, \PR{71}, 38 (1947).
\ref [2] M. Maggiore, \NP{B647}, 69 (2002).
S. Hossenfelder, \PR{D75}, 105005 (2007); \arx{1403.2080} (2014).
\ref [3] S. Benczik, L.N. Chang, D. Minic, N. Okamura, S. Rayyan and T. Takeuchi,
\PR{D66}, 026003 (2002).
C. Leiva, J. Saavedra and J R Villanueva, Pramana - J. Phys. {\bf 80},
945 (2013).
B. Iveti\'c, S. Meljanac and S. Mignemi, \arx{1307.7076} (2013).
\ref [4]
R. Banerjee, S. Kulkarni and S. Samanta, \JHEP{0605}, 077 (2006).
S. Mignemi, \PL{B672}, 186 (2009).
\ref [5] E. J. Hellund and K. Tanaka \PR{94}, 192 (1954).
G. Jaroszkiewicz, \JoP{A28}, L343 (1995).
J.M. Romero and A. Zamora, \PR{D70}, 105006 (2004); \PL{B661}, 11 (2008).
E.R. Livine and D. Oriti, \JHEP{0506}, 050 (2005);
F. Girelli and E. Livine, \JHEP{1103}, 132 (2011).
M.V. Battisti and S. Meljanac, \PR{D79}, 067505 (2009); \PR{D82}, 024028 (2010).
S. Mignemi, \PR{D84}, 025021 (2011).
Lei Lu and A. Stern, \NP{B854}, 894 (2011); \NP{B860}, 186 (2012).
\ref [6]
C.N. Yang, \PR{72}, 874 (1947).
J. Kowalski-Glikman and L. Smolin, \PR{D70}, 065020 (2004).
C. Chryssomakolos and E. Okon, \IJMP{D13}, 1817 (2004).
H.G. Guo, C.G. Huang and H.T. Wu, \PL{B663}, 270 (2008).
S. Mignemi, \CQG{26}, 245020 (2009); \CQG{29}, 215019 (2012).
M.C. Carrisi and S. Mignemi, \PR{D82}, 105031 (2010).
R. Banerjee, K. Kumar and D. Roychowdhury, \JHEP{1103}, 060 (2011).
\ref [7] S. Mignemi, \arx{1308.0673} (2013).
\ref [8] C.M. Will, Living Rev. Rel. {\bf 9}, 3 (2005).

\endref
\end